\documentclass[10pt,conference]{IEEEtran}
 
\usepackage{soul} 
\usepackage{color} 
\sethlcolor{white}

\usepackage{cite}
\ifCLASSINFOpdf
  \usepackage[pdftex]{graphicx}
  \graphicspath{{../figures/}}
\fi

\usepackage{array}
\usepackage[caption=false,font=footnotesize]{subfig}
\usepackage{url}
\usepackage{courier}
\usepackage{stfloats}
\usepackage{tablefootnote}
\usepackage{booktabs}
\usepackage{caption}
\usepackage{varwidth}
\usepackage{microtype}
\DeclareCaptionFont{ls}{\lsstyle}
\DeclareCaptionFormat{myformat}{%
  \begin{varwidth}{\linewidth}%
    \centering
    #1#2#3%
  \end{varwidth}%
}

\usepackage[T1]{fontenc}
            
\newcommand{\inlinecode}[1]{{\small \texttt {#1}}}

\newcommand{\doxygen}{doxygen}
\newcommand{\api}{API}
\newcommand{\docio}{our tool}
\newcommand{\funcwatch}{funcWatch}
\newcommand{\exselect}{ioSelect}
\newcommand{\expresent}{ioPresent}
\newcommand{\ffmpeg}{ffmpeg}
\newcommand{\libssh}{libssh}
\newcommand{\protobuf}{protobuf-c}
\newcommand{\io}{I/O}

\begin{document}
\fnbelowfloat
\captionsetup[table]{format=myformat, labelsep=newline}

\title{Documenting API Input/Output Examples}
\author{\IEEEauthorblockN{Siyuan Jiang, Ameer Armaly, Collin McMillan, Qiyu Zhi, and Ronald Metoyer}
\IEEEauthorblockA{Department of Computer Science and Engineering\\
University of Notre Dame, Notre Dame, IN, USA\\
\{sjiang1, aarmaly, cmc, qzhi, rmetoyer\}@nd.edu}}
\maketitle

\begin{abstract}
When learning to use an Application Programming Interface ({\api}), programmers need to understand the inputs and outputs ({\io}) of the {\api} functions. Current documentation tools automatically document the static information of {\io}, such as parameter types and names. What is missing from these tools is dynamic information, such as {\io} examples{\textemdash}actual valid values of inputs that produce certain outputs. 

In this paper, we demonstrate a \emph{prototype} toolset we built to generate {\io} examples. {\docio} logs {\io} values when {\api} functions are executed, for example in running test suites. Then, {\docio} puts {\io} values into {\api} documents as {\io} examples. 

{\docio} has three programs: 1) {\funcwatch}, which collects {\io} values when {\api} developers run test suites, 2) {\exselect}, which selects one {\io} example from a set of {\io} values, and 3) {\expresent}, which embeds the {\io} examples into documents. In a preliminary evaluation, we used {\docio} to generate four hundred {\io} examples for three C libraries: {\ffmpeg}, {\libssh}, and {\protobuf}.
\end{abstract}

\vspace{-0.1cm}
\section{Introduction}
Recent studies by Duala-Ekoko and Robillard~\cite{Duala-Ekoko2012} and by Sillito~\emph{et al.}~\cite{Sillito2008} found that when learning Application Programming Interfaces ({\api}), programmers look for input/output examples, \textbf{actual values} of a function's input/output, in order to understand the function. In particular, the programmers asked ``How does this data structure look at runtime?''~\cite{Sillito2008} and ``we have a newInstance(String) method that takes a String argument and I have no idea what this String is supposed to be''~\cite{Duala-Ekoko2012}.

Programmers often refer to {\api} documents to answer such questions~\cite{Forward2002,Robillard2010}. However, a study by Maalej and Robillard shows that most of the {\api} documents do not contain input/output ({\io}) examples~\cite{Maalej2013}. One reason for the documents lacking {\io} examples is that {\api} developers have little tool support for documenting {\io} examples. Current documentation tools, such as {\doxygen}~\cite{Doxygen}, JavaDoc~\cite{Javadoc}, automate the process of converting comments and source code into structured html pages. If {\api} developers write {\io} examples in comments, the tools can present the {\io} examples in html pages. Otherwise, the tools have no {\io} examples to present.


When there is little help from {\api} documents, programmers (the users of {\api} libraries) often try a variety of approaches to get input/output examples~\cite{Ko2006,Lawrance2013,Roehm2012}. For instance, programmers 1) search or ask a question online (e.g., \url{stackoverflow.com}), 2) look for existing code that use the {\api} (e.g., open source repositories), or 3) write code to do experiments about the {\api}. If the {\api} documents have input/output examples, programmers may not need these approaches. 

To help {\api} developers add {\io} examples into {\api} documents, we built {\docio}, which is a prototype toolset that (1) logs actual {\io} values when {\api} functions are executed, (2) selects an example from the logged values for each {\api} function, and (3) visualizes the examples to {\api} document. 

In the rest of the paper, we will describe what {\docio} adds to {\api} documents, how {\api} developers can use {\docio}, how {\docio} creates input/output examples, the preliminary results, the related work, and the future work. 

\section{{\docio} in a Nutshell}
\label{sec:nutshell}
\label{sec:onlineappendix}
{\docio} is for developers of C {\api} libraries to create {\api} documents with {\io} examples{\textemdash}actual values of input/output{\textemdash}of {\api} functions. {\docio} has three programs: {\funcwatch}, {\exselect} and {\expresent}. First, {\funcwatch} logs {\io} values of a function when the function is executed (in tests or in other executable programs). Often the function is executed multiple times. Then, {\exselect} selects one of the calls. Finally, {\expresent} adds the {\io} values of the selected call into the {\api} document. IoPresent is a patch for the documentation tool, {\doxygen}~\cite{Doxygen}. The patch enables {\doxygen} to visualize {\io} examples into html pages.

This website contains a virtual machine that is set up for running {\docio}, the videos about {\docio}, the installation guides, and three {\api} documents generated by {\docio}.

\section{Motivational Example}
\label{sec:motivation}
\begin{figure}[tb]
\centering
\includegraphics[width=3.5in]{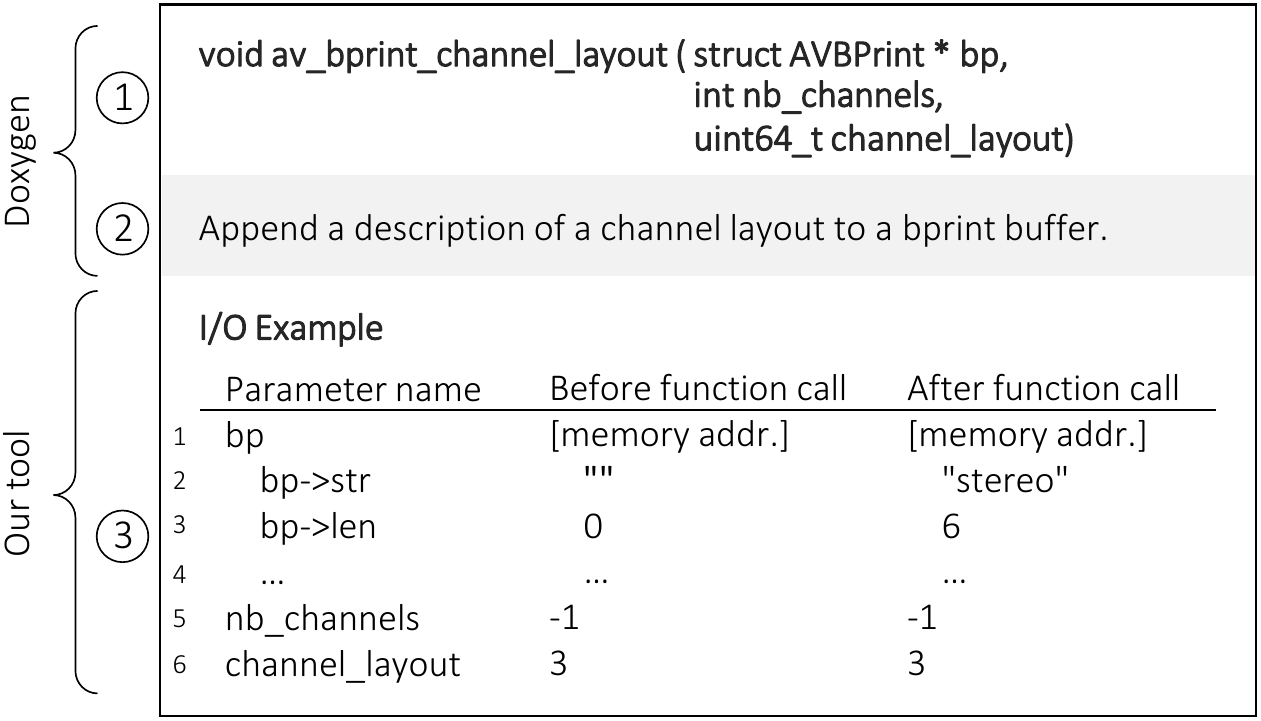}
\vspace*{-0.3cm}
\caption{\footnotesize{The document of function \texttt{av\_bprint\_channel\_layout} in {\ffmpeg}. The document has three parts. Part 1 is the function declaration. Part 2 is a description of the function. Parts 1 and 2 are generated by {\doxygen}. Part 3 is an input/output example created by {\docio}. The input/output example is a table with three columns: ``Parameter name'', ``Before function call'', and ``After function call''. The first column lists the names of parameters. The second column lists the values of the parameters before \texttt{av\_bprint\_channel\_layout} is called, i.e. input values. The third column lists the values of the parameters after \texttt{av\_bprint\_channel\_layout} is called, i.e. output values. \\[3pt]
Each row in the table represents a pair of input/output values. For example, the first row represents the values of the first parameter, \texttt{bp}, which is pointer-type. Because the actual values of pointers---memory addresses---are not meaningful, {\docio} does not put actual values in the table but ``[memory addr.]'' to indicate the values are memory addresses. \\[3pt]
The value that \texttt{bp} points to is a struct variable, which is composed of the fields of the struct. The second, third and fourth rows represent the fields. The three dots in the fourth row does not represent a field but represent three more fields of the struct variable. {\docio} presents the three fields in the actual document, but we omit them due to page limits. Section~\ref{sec:motivation} discussed this figure in detail.}}
\label{fig:motivation}
\vspace*{-0.1cm}
\end{figure}

\def\UrlFont{\rmfamily}
In this section, we will show what {\docio} adds to {\api} documents. Figure~\ref{fig:motivation} shows a document generated by {\docio} and {\doxygen}. The document is about the function \inlinecode{av\_bprint\_channel\_layout} in {\ffmpeg}.  

The {\api} document is composed of three parts. Part 1 is the declaration of \inlinecode{av\_bprint\_channel\_layout}. Part 2 is a sentence describing the function. Parts 1 and 2 are generated by {\doxygen}. Part 3 is what {\docio} adds to the document, which is an {\io} example. In the rest of the section, we will walk through the three parts of the document from the perspective of a user who wants to learn \inlinecode{av\_bprint\_channel\_layout}. 

First, the user reads Part 1 and gets some basic information: (1) the name of the function, (2) the return type, which is \inlinecode{void}, and (3) the types and the names of the parameters. According to the previous studies~\cite{Duala-Ekoko2012,Sillito2008}, the user may have the following questions after reading Part 1:
\\[5pt]
\indent Q1. What does \inlinecode{av\_bprint\_channel\_layout} do? \\
\indent Q2. What is \inlinecode{struct AVBPrint}? \\
\indent Q3. What is \inlinecode{nb\_channels}?
\\[5pt]
\indent Second, the user reads Part 2, which answers Q1: the function ``append a description of a channel layout to a bprint buffer.'' From this description, the user may expect an input representing ``a description of a channel layout'', and an output representing ``a bprint buffer''. The names of the parameters indicate that the parameter \inlinecode{channel\_layout} is a channel layout and \inlinecode{bp} is a bprint. But the user still does not know where the description of the channel layout is, and where the buffer of the bprint is. In other words, the user may have the questions:
\\[5pt]
\indent Q4. Where is the description of a channel layout?\\
\indent Q5. Where is the buffer of the bprint?
\\[5pt]
\indent Part 3, the {\io} example, provides some hints for some unanswered questions. The {\io} example shows that \inlinecode{bp-{\textgreater}str} changed from an empty string to ``stereo''. So the reader can guess that \inlinecode{bp-{\textgreater}str} is the buffer of the bprint (answers Q5). 

The {\io} example also shows that \inlinecode{nb\_channels} can be ``-1'', which is related to Q3. For Q4, the user still does not know where the description of a channel layout is. What the reader can know from the {\io} example is that when channel\_layout is three, the description is ``stereo''.

In summary, all the three parts have useful information that helps the user to understand \inlinecode{av\_bprint\_channel\_layout}. Part 1 has static information extracted from source code. Part 2 has the high-level description that only {\api} developers can provide. {\docio} adds Part 3, the actual, concrete input/output values of a function. 

\section{Usage of {\docio}}
The usage of {\docio} has three steps: 1) run {\funcwatch}, 2) run {\exselect}, and 3) run {\expresent}. If an {\api} developer wants to get {\io} examples for all the functions in the library, the developer needs to have a list of the function names for running {\funcwatch} (so that {\funcwatch} knows which function to log). To help developers with this situation, we modified {\doxygen} so that {\doxygen} writes all the function names in a file when {\doxygen} is executed. 

We put a demo video in our online appendix (Section~\ref{sec:onlineappendix}) in which we followed these steps and generated nearly two hundred {\io} examples for the {\ffmpeg} library.

\section{Envisioned Users}
The envisioned users of {\docio} are the developers of {\api} libraries. The developers can use {\docio} to create {\api} documents with {\io} examples. Although the developers are the envisioned users of {\docio}, the {\api} users (the programmers who use and learn {\api} libraries) are the ones that read the documents and benefit from the {\io} examples.

Additionally, the programs in {\docio} can be used independently for other purposes. For example, {\funcwatch} can be used to detect whether a parameter always takes the same value.

\section{Design of {\docio}}
\begin{figure}[tb]
\centering
\includegraphics[width=3.35in]{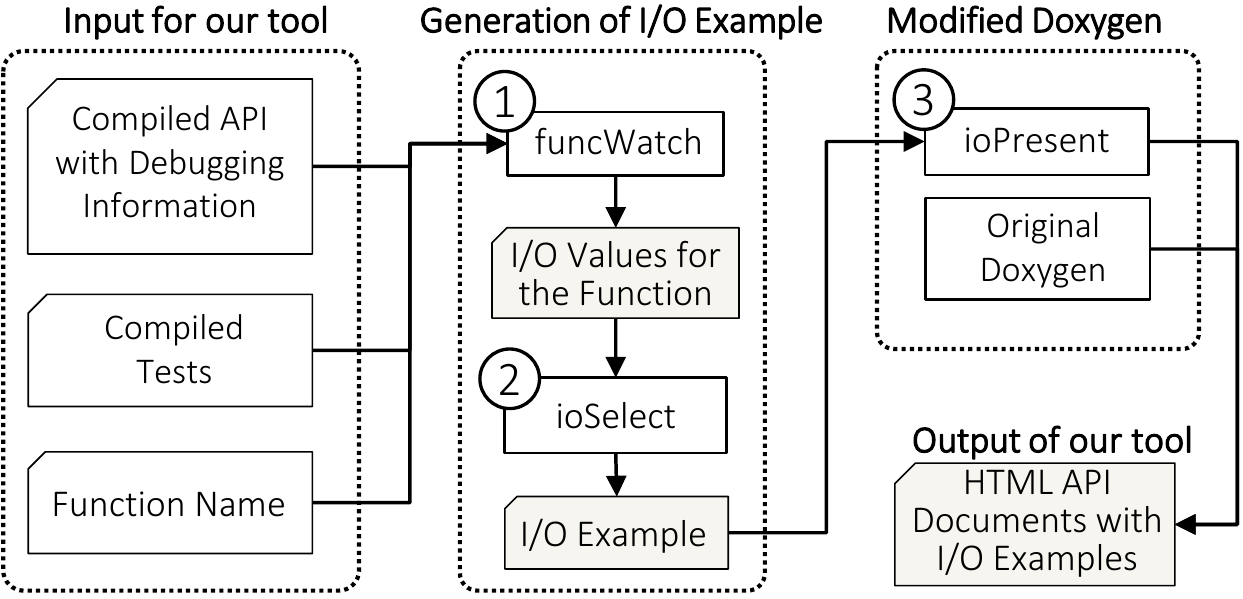}
\caption{\footnotesize{The architecture of {\docio}. The inputs of {\docio} are (1) compiled {\api}, (2) compiled {\api} tests, (3) an {\api} function name. The three programs in {\docio} are labeled with numbers: (1) {\funcwatch}, (2) {\exselect}, and (3) {\expresent}. FuncWatch runs tests and code examples, and logs input/output values of the function specified by the input. IoSelect chooses an example from all the logged {\io} values. IoPresent adds the example to html documents.}}
\label{fig:architecture}
\vspace*{-0.4cm}
\end{figure}

In this section, we will describe the design of {\docio}. Figure~\ref{fig:architecture} shows the architecture of {\docio}, which illustrates the relationship among the three programs: {\funcwatch}, {\exselect}, and {\expresent}. In the following subsections, we will describe how we design and implement each program.

\subsection{FuncWatch: Logging I/O values}
\label{sec:tool2}
The input of {\funcwatch} is an {\api} function name and a test. FuncWatch runs and monitors the test, and, during the execution of the test, {\funcwatch} logs the {\io} values of the given {\api} function.

We built {\funcwatch} based on the code of Flashback~\cite{Armaly2016}, which uses libdwarf~\cite{libdwarf}, a library that provides access to Dwarf debugging information. By using libdwarf, {\funcwatch} inserts breakpoints at the entry and the exits of the given function (a C function may have multiple exits). 

Then, {\funcwatch} runs the given test with the breakpoints. After a breakpoint is reached and the test paused, {\funcwatch} collects the parameters/return values of the {\api} function. Furthermore, {\funcwatch} is able to dereference pointers at runtime, so {\funcwatch} can collect not only pointers' values, but also the values that the pointers point to. FuncWatch also calculates the addresses of a struct variable's fields, so that {\funcwatch} can log the values of the fields. Similarly, {\funcwatch} also work for union and enum types.

In C, arrays are often passed without size information to functions. While a common practice is to pass the size of an array as a separate parameter, the naming conventions differ across different projects and developers. So there is no simple way to automatically detect the size parameters. In this case, {\funcwatch} logs only the first item of an array.

In a test, the given function can be called multiple times. FuncWatch logs a call id for every parameter/return value in order to distinguish the values collected from different calls.

\subsection{{\exselect}: Choosing I/O examples}
\label{sec:tool3}
The input of {\exselect} is a set of {\io} values of an {\api} function. Each value is labeled with a call id logged by {\funcwatch}. IoSelect selects a call id randomly, and all the values with the same call id become the {\io} example for the function. Because {\exselect} selects {\io} examples randomly, for the same {\api} function and the same test, {\docio} may create different {\io} examples. We implemented {\exselect} as a standalone program so that we can easily add new algorithms for selecting {\io} examples in the future.

\subsection{{\expresent}: Adding I/O examples to documents}
\label{sec:tool4}
The input of {\expresent} is a set of {\io} examples. Each {\api} function has only one {\io} example. The output of {\expresent} is html {\api} documents with the {\io} examples.

The main functionality of {\expresent} is to visualize an {\io} example into a html table like the table in Figure~\ref{fig:motivation}. For example, in Figure~\ref{fig:motivation}, {\expresent} automatically detects \inlinecode{bp->str} is a field of a struct variable that \inlinecode{bp} points to. Then, {\expresent} automatically indents the row of \inlinecode{bp->str} and makes the row of \inlinecode{bp} collapsible (so that all the values that \inlinecode{bp} points to can be hidden if needed).

We implemented {\expresent} as a patch of {\doxygen}. Specifically, we modified two parts of {\doxygen}: \textit{Data Organizer} and \textit{Html Generator}. 

\textit{Data Organizer} builds data structures to represent elements in the {\api} libraries (that are going to be documented), such as classes and functions. First, we added a new class to represent a new type of element---the {\io} examples. Second, we added {\io} examples into the data structure that represents a function. We also modified \textit{Html Generator}, in which we implemented the process of visualizing {\io} examples in html format.

\section{Preliminary Evaluation}
To assess the ability of {\docio} to generate {\io} examples for real-world libraries, we did a preliminary evaluation to answer the question: how many functions will have {\io} examples by using {\docio}?

In this evaluation, we used {\docio} on three {\api} libraries: {\ffmpeg}, {\libssh}, and {\protobuf}. For each library, we executed {\funcwatch} on every function (found by {\doxygen}) and every test in the test suites. Then, we executed {\exselect} and {\expresent} to generate {\api} documents with {\io} examples. 

In total, {\docio} generates {\io} examples for 402 functions, 191 functions in {\ffmpeg}, 202 in {\libssh}, and 9 in {\protobuf}. To put the numbers in context, Table~\ref{tab:coverage} shows the number of functions in source code, the number of functions in {\api} documents, along with the number of functions that have {\io} examples. The {\api} documents have much fewer functions than source code, because {\doxygen} includes only the functions that are commented in a predefined format. This default setting helps {\doxygen} exclude the internal functions, which should not be seen by {\api} users.

\begin{table}[tb]
\centering
\caption{\sc{The number of functions that \newline have {\io} examples generated by {\docio}}}
\label{tab:coverage}
\begin{tabular}{lrrr}
\toprule
{\api} library 
		  & \begin{tabular}{@{}c@{}}\# of functions\\in source code\end{tabular} 
                  & \begin{tabular}{@{}c@{}}\# of functions\\in api doc. \end{tabular} 
                  & \begin{tabular}{@{}c@{}}\# of functions\\with {\io} examples  \\\end{tabular} \\
\midrule
{\ffmpeg}     &                      20,347 &                                       625 &                                      191 \\
{\libssh}        &                        1,473 &                                       519 &                                      202 \\
{\protobuf}   &                        3,171 &                                         18 &                                         9 \\
\bottomrule
\end{tabular}
\end{table}

{\docio} did not generate {\io} examples for all the functions that are in {\api} documents, because not all the functions were executed in the test suites. For example, {\ffmpeg} has more than twenty thousand functions defined in source code. Only about six hundred functions are documented. {\docio} created {\io} examples for 30\% of the documented functions.

\section{Visualizing all {\io} values}
\begin{figure*}
\centering
\includegraphics[width=6.3in]{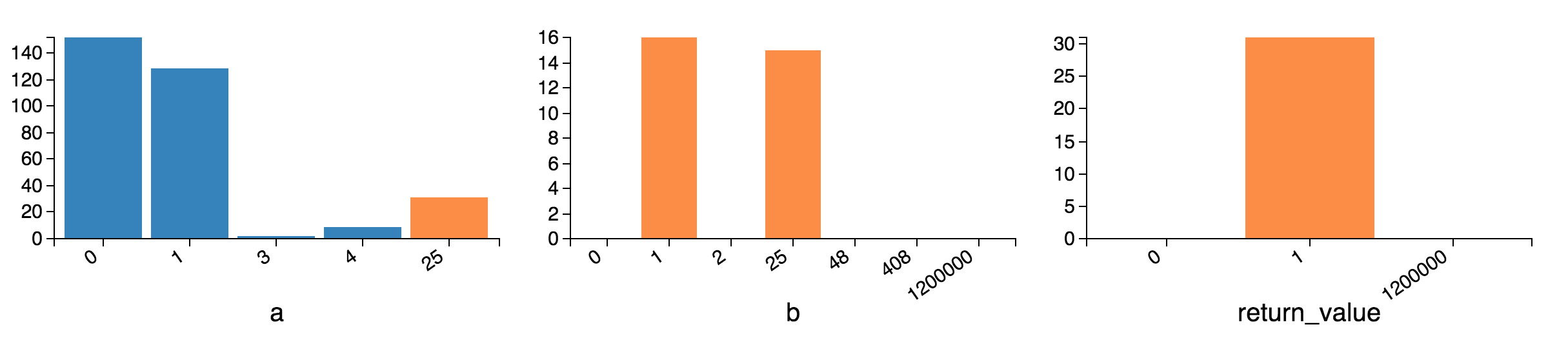}
\vspace*{-0.2cm}
\caption{\footnotesize{All the logged {\io} values for function \texttt{av\_gcd}. Three bar charts correspond to three variables: \texttt{a}, \texttt{b}, and \texttt{return}. In this figure, the user has moused over the possible value ``25'' in \texttt{a}, and the corresponding values for \texttt{b} and \texttt{return} are updated and highlighted.}}
\vspace*{-0.3cm}
\label{fig:allvalues}
\end{figure*}

Sometimes, it is useful to see all the possible values of the parameters and return variables of an {\api} function. So we built a prototype to visualize all the values logged by {\funcwatch} for an {\api} function. For each parameter in a function, we draw a bar chart of all the possible values of the parameter, where the height of each bar represents the frequency of that value. For example, in Figure~\ref{fig:allvalues}, function \inlinecode{av\_gcd} has three bar charts. Each bar chart represents a parameter or a return variable. Parameter \inlinecode{a} has five different values: \inlinecode{0, 1, 3, 4, 25} (logged by {\funcwatch}). The most frequent value for \inlinecode{a} is \inlinecode{0}, which occurs in more than 140 calls.

The bar charts are interactive to show the relationships between the parameters and return variables. For example, in Figure~\ref{fig:allvalues}, the cursor is hovered over bar \inlinecode{25} in bar chart \inlinecode{a}. The other two bar charts are updated to reflect the values collected when \inlinecode{a} was \inlinecode{25}. In other words, Figure~\ref{fig:allvalues} shows that there are about 30 calls where \inlinecode{a} is \inlinecode{25}. In these calls, \inlinecode{b} is either \inlinecode{1} or \inlinecode{25}. And all these calls return \inlinecode{1}. 

\section{Related Work}
In this section, we will present two categories of related work. \textbf{The first category} is about introducing different types of contents into {\api} documents, such as information from StackOverflow posts~\cite{Treude2012}, usage information~\cite{Stylos2009}, frequently-asked-questions from the Web~\cite{Chen2014}, and code example summaries mined from the Web~\cite{Kim2013,Subramanian2014}. Different from these approaches, {\docio} does not mine the Web, and {\docio} adds {\io} values instead of source code examples into {\api} documents.

\textbf{The second category} is about collecting and analyzing dynamic information, which can be useful in {\api} documentation. For example, \textit{Daikon} analyzes dynamic information to report likely program invariants~\cite{Ernst2007}. The reports can be useful in {\api} documents. We plan to embed this type of reports to {\api} documents in the future in order to study whether and how the reports can help programmers use {\api} functions.

\section{Future Work and Limitations}
{\docio} is a prototype toolset, which serves as our first step towards investigating what information is useful for {\api} users. Previous studies have shown that programmers ask questions about dynamic information~\cite{Duala-Ekoko2012,Sillito2008}, such as {\io} examples. What is not shown is whether and how this information helps programmers. First, we will conduct an empirical study to assess the usefulness of {\io} examples. In this study, we plan to hire programmers to complete certain program tasks with and without {\io} examples in documents. By comparing the programmers' performance with and without {\io} examples, we can see how {\io} examples can help programmers.

Furthermore, we plan to design and implement different selection strategies for {\exselect}. Our assumption is that some {\io} examples are more useful than others. We plan to investigate some potential characteristics of useful {\io} examples. For example, the most common {\io} values may be more useful.

Additionally, the quality and the quantity of the {\io} examples generated by {\docio} depend on the test suites of the {\api} libraries. In the future work, we may consider code examples, tutorial examples, and real-world applications that use the {\api} libraries in addition to the test suites to generate {\io} examples.

Implementation-wise, {\docio} currently has three major limitations. First, {\funcwatch} does not support 64-bit memory addressing so {\docio} can work only in 32-bit operating systems. Second, {\funcwatch} collects only the values for statically-linked libraries. Third, {\funcwatch} collects all the {\io} values no matter whether a test passes or fails. If a test fails, the logged values may be invalid. The current workaround for this limitation is to detect the outputs of the tests in a script and discard the outputs of {\funcwatch} for those failed tests. 

\section{Conclusion}
To conclude, we built and demonstrated a prototype toolset, {\docio}. The goal of {\docio} is to help {\api} developers create and add {\io} examples into {\api} documents. {\docio} has three programs: {\funcwatch}, {\exselect}, and {\expresent}. Our preliminary evaluation shows that {\docio} can generate four hundred {\io} examples for three real-world C libraries.


\section*{Acknowledgment}
We thank Douglas Smith for his work in visualization of all {\io} values. This work was partially supported by the NSF CCF-1452959 and CNS-1510329 grants, and the Office of Naval Research grant N000141410037. Any opinions, findings, and conclusions expressed herein are the authors' and do not necessarily reflect those of the sponsors.

\bibliographystyle{IEEEtran}
\bibliography{IEEEabrv,biblio}

\end{document}